\documentclass[aip,amsmath,amssymb,reprint,times,superscriptaddress,longbibliography]{revtex4-1}

\usepackage{float}
\usepackage{graphicx}
\usepackage{dcolumn}
\usepackage{bm}
\usepackage{mathrsfs,amssymb}
\usepackage{ulem}
\usepackage{xcolor}
\usepackage{soul}
\usepackage[colorlinks=true,linkcolor=blue,citecolor=blue,anchorcolor=blue,urlcolor=blue]{hyperref}
\usepackage{lineno}

\expandafter\ifx\csname package@font\endcsname\relax\else
 \expandafter\expandafter
 \expandafter\usepackage
 \expandafter\expandafter
 \expandafter{\csname package@font\endcsname}%
\fi
\begin{document}

\title{Voltage-driven charge-mediated fast 180 degree magnetization switching in nanoheterostructure at room temperature}

\author{Min Yi}
\email{yi@mfm.tu-darmstadt.de}
\author{Hongbin Zhang}
\affiliation{Institute of Materials Science, Technische Universit\"{a}t Darmstadt, 64287 Darmstadt, Germany}
\author{Bai-Xiang Xu}
\affiliation{Institute of Materials Science, Technische Universit\"{a}t Darmstadt, 64287 Darmstadt, Germany}

\date{\today}

\begin{abstract}
Voltage-driven 180$^\circ$ magnetization switching without electric current provides the possibility for revolutionizing the spintronics. We demonstrated the voltage-driven charge-mediated 180$^\circ$ magnetization switching at room temperature by combining first-principles calculations and temperature-dependent magnetization dynamics simulation. The electric field ($E$) induced interface charge is found to allow a giant modulation of the magnetic anisotropy ($K$) of the nanomagnet. Particularly $K$ is revealed to vary linearly with respect to $E$ and the epitaxial strain. Magnetization dynamics simulations using the so-obtained $K$ show that both in-plane and perpendicular 180$^\circ$ switching can be achieved by $E$ pulses. The temperature effect renders the 180$^\circ$ switching as probability events. Statistical analysis indicates a fast (around 4 ns) and low-error-probability 180$^\circ$ switching achievable at room temperature by controlling the magnitude of $E$ and the pulse width. The study inspires the rational design of miniaturized nanoscale spintronic devices where thermal fluctuation has a great impact.
\end{abstract}

\maketitle

\section*{Introduction}
Recently, the voltage control of magnetism without electric current has been extensively studied in order to achieve minimum power consumption and device miniaturization in next-generation of spintronics.\cite{1,2,3,4,5,6,7,8,9,v0} Such a control is usually implemented through the magnetoelectric (ME) coupling in heterostructures which possess coupled magnetic and electric properties. Generally, in ME heterostructures voltage can control the magnetism through the interfacial mechanisms such as elastic coupling via strain transfer,\cite{10,11,12,13,14,15,16,17,18,19,20,21,22,23,24,25,26,27,28,29,v1,v2} charge modulation,\cite{30,31,32,33,34,35,36a,36,37,38,39,v1,v2} interface bonding,\cite{40,41,42,43,44} and exchange coupling.\cite{28,45,46,47,48} For the ferromagnetic/ferroelectric heterostructures, elastic coupling mechanism is most extensively investigated, i.e. a strain generated in a ferroelectric layer by a voltage is transferred to the ferromagnetic layer through the interface and thus can be used to tailor magnetization through the magnetoelastic coupling. The elastic coupling mechanism can work at the bulk level. All the other three mechanisms are localized to the region near the interface.
Among them, the exchange coupling is localized, but can generate exchange spring effects to affect the magnetization in the bulk.{\cite{48a}} Compared to the elastic coupling mechanism, they offer more freedom to design reasonable and optimum nanoscale devices in which the interface plays a critical role, owing to the diversity and tunability of the interfacial structure and chemistry.

By using these various mechanisms, a voltage induced 180$^\circ$ magnetization switching is of great interests. For example, in order to achieve a high signal-to-noise ratio in magnetic tunnel junction (MTJ), a significantly large electric resistance change of MTJ is required, which can only be achieved by a 180$^\circ$ switching in the free layer of MTJ. Trailblazing experiments have demonstrated strain-mediated and exchange-coupling-mediated voltage-driven/assisted magnetization switching in Ni/BaTiO$_3$,\cite{22}, Co/PMN-PT,\cite{39} and CoFe/BiFeO$_3$\cite{46,47} heterostructures, respectively.

Nevertheless, the voltage-driven full 180$^\circ$ switching is not achieved in some experiments. For example, in the CoFe/BiFeO$_3$ only a net magnetization reversal is realized.{\cite{46}} The electric field is found to decrease the coercivity, but the switching is still achieved by magnetic field.{\cite{36,47}} Also some forms of magnetic fields are involved in experiments.{\cite{39,46,22}}
From the theoretical point, a large number of studies are devoted to the strain-mediated 180$^\circ$ switching either by designing the shape of magnets\cite{29} or by using the 3D precessional switching dynamics.\cite{12,13,14,17,18,23,24,26}

Alternatively, interface charge modulation has been deemed as an ideal way for realizing magnetic switching in thin film heterostructures.\cite{30,31,32,33,34,35,36a,36,37,38,39} But plenty of experimental and theoretical investigations show charge-modulated coercivity or charge-mediated magnetization switching between two states which are not exactly rotated by 180$^\circ$.\cite{30,31,36a,36,34,37}
For example, an electric field induced change in the number of unpaired $d$ electrons in FePt and FePd can induce their coercivity change by several percent, but the voltage-driven magnetization switching is not addressed.{\cite{30}} Controlling the charge carrier concentration in (Ga,Mn)As by an electric field can manipulate the magnetization direction by only 10$^\circ$.{\cite{36a}} In Fe/MgO system, through the voltage control of relative occupation of $3d$ orbitals in interfacial Fe atoms, an electric field of 0.1 V/nm is shown to cause a 40$\%$ change in the magnetic anisotropy of Fe.{\cite{34}} In CoFeB/MgO/CoFeB, electric field is shown to tune the coercivity for magnetization switching driven by a magnetic field.{\cite{36}} Coherent control of magnetization using voltage signals is realized in FeCo/MgO with an additional magnetic field of 70 mT and switching angle less than 180$^\circ$.{\cite{37}}
Meanwhile, most studies deal with the thin film structure with a lateral size of several hundred nanometers to several micrometers or with a thickness of several tens nanometers. In such cases, the magnet volume is relatively large so that the temperature effect on the switching dynamics is often ignored,\cite{10,12,13,14,17,18,19,24,29,31,48} or shown to be not so strong.\cite{23,26} However, in terms of the device miniaturization, if spintronic devices are further scaled down to nanodot shape, i.e. with the size of several tens of nanometers in the lateral direction, the huge reduction of the magnet volume will intensify the effects of temperature-induced thermal fluctuations. For instance, the granular film with $L1_0$-ordered FePt epitaxially grown on MgO substrate, which has been considered as a promising candidate for memory devices, usually contains FePt nanograins with a size of several nanometers to several tens of nanometers. If such a small-scaled FePt/MgO heterostructure is utilized, considering the effect of thermal fluctuations on the magnetization dynamics is indispensable.

\begin{figure}[!t]
\centering
\includegraphics[width=8.6cm]{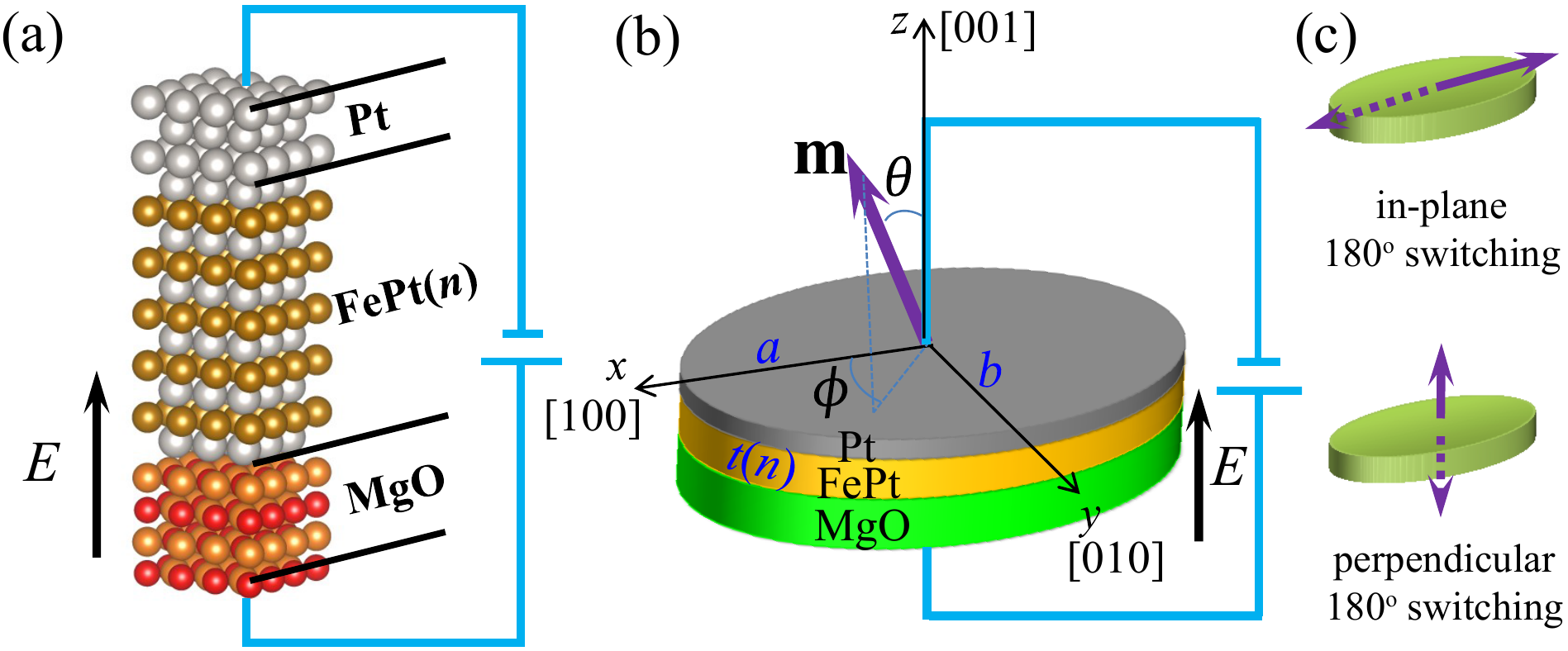}
\caption{Material model system for (a) first-principles calculations and (b) magnetization dynamics simulation. (c) Illustrations of in-plane and perpendicular 180$^\circ$ switching.}
\label{f1}
\end{figure}

In this work, we take epitaxial Pt/FePt/MgO heterostructures as a model system with a lateral size of several tens of nanometers. 
An elliptical shape of FePt (Fig. {\ref{f1}}(b)) is chosen, which allows the magnetic anisotropy within the $x$-$y$ plane and thus the in-plane magnetization switching. The electric field induced change in the FePt/MgO interface charge can tailor the magnetocrystalline anisotropy energy (MAE) of FePt layer, thus manipulating the magnetic easy axis of the heterostructure system. Meanwhile, the epitaxial strain plays a role in determining whether the equilibrium magnetic state without electric field is in-plane or perpendicular. By controlling the electric field, epitaxial strain and magnetization dynamics, the in-plane and perpendicular 180$^\circ$ magnetization switching (Fig. {\ref{f1}}(c)) is achievable in the case of in-plane and perpendicular equilibrium magnetic state, respectively.
Specifically, combining first-principles calculation and temperature-dependent magnetization dynamics, we demonstrate the in-plane and perpendicular 180$^\circ$ magnetization switching at room temperature.
It is anticipated that the present study provides valuable insight into the design of voltage control of both in-plane and perpendicular 180$^\circ$ switching at room temperature for achieving low-power, high-speed, non-volatile, and highly compact spintronics.

\section*{Results and discussion}{\label{s2}}
The material model system for first-principles calculations is shown in Fig. \ref{f1}(a). Since the lattice parameter of bulk MgO is $\sim 4.22$ \AA ~ and that of FePt is $\sim 3.86$ \AA, the lattice mismatch is up to $\sim 8.5\%$. In experiments, MgO with tunable lattice strain can be epitaxially grown at different temperature on different substrates like Cu,\cite{50}, Ge,\cite{51} Si,\cite{52} etc. So different in-plane epitaxial strains ($\varepsilon_\text{MgO}$) relative to the equilibrium lattice parameter in MgO at different FePt layer number ($n$) are adopted to study their effects. The material system for micromagnetic dynamics analysis is shown in Fig. \ref{f1}(b). The FePt nanomagnet is an elliptical cylinder with height $t(n)$, semimajor axis $a$, and semiminor axis $b$. Two angles are used to describe the magnetization state, as presented in the Methods section.

\begin{figure}[!t]
\centering
\includegraphics[width=8.6cm]{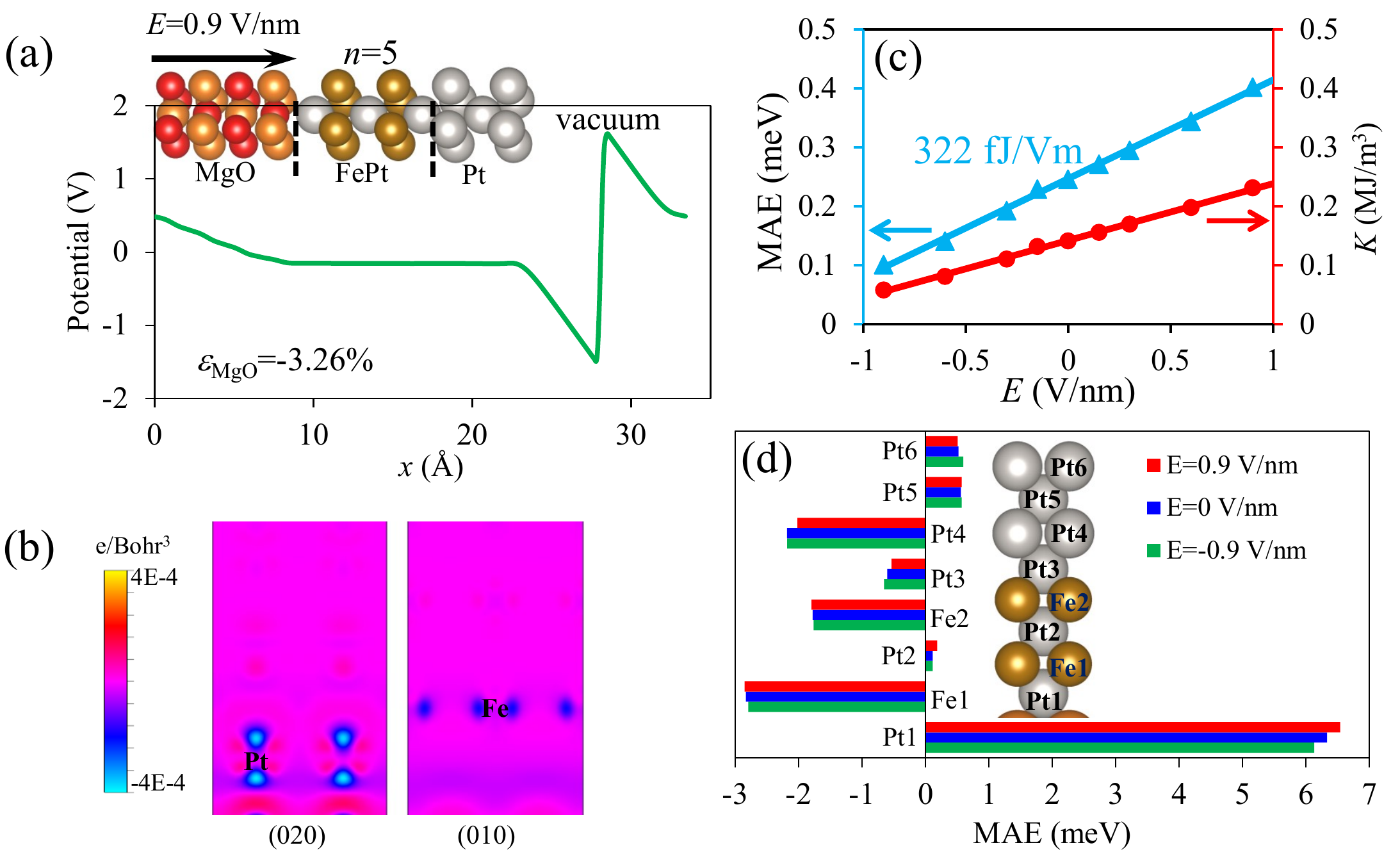}
\caption{Typical first-principles results for $n=5$ and $\varepsilon_\text{MgO}=-3.26\%$. (a) Electrostatic potential distribution. (b) Electric field ($E=$0.9 V/nm) induced charge density change near Pt and Fe atoms close to the MgO/FePt interface. (c) MAE as a function of the external electric field. (d) Atom-resolved MAE.}
\label{f2}
\end{figure}

\begin{figure}[!t]
\centering
\includegraphics[width=8.6cm]{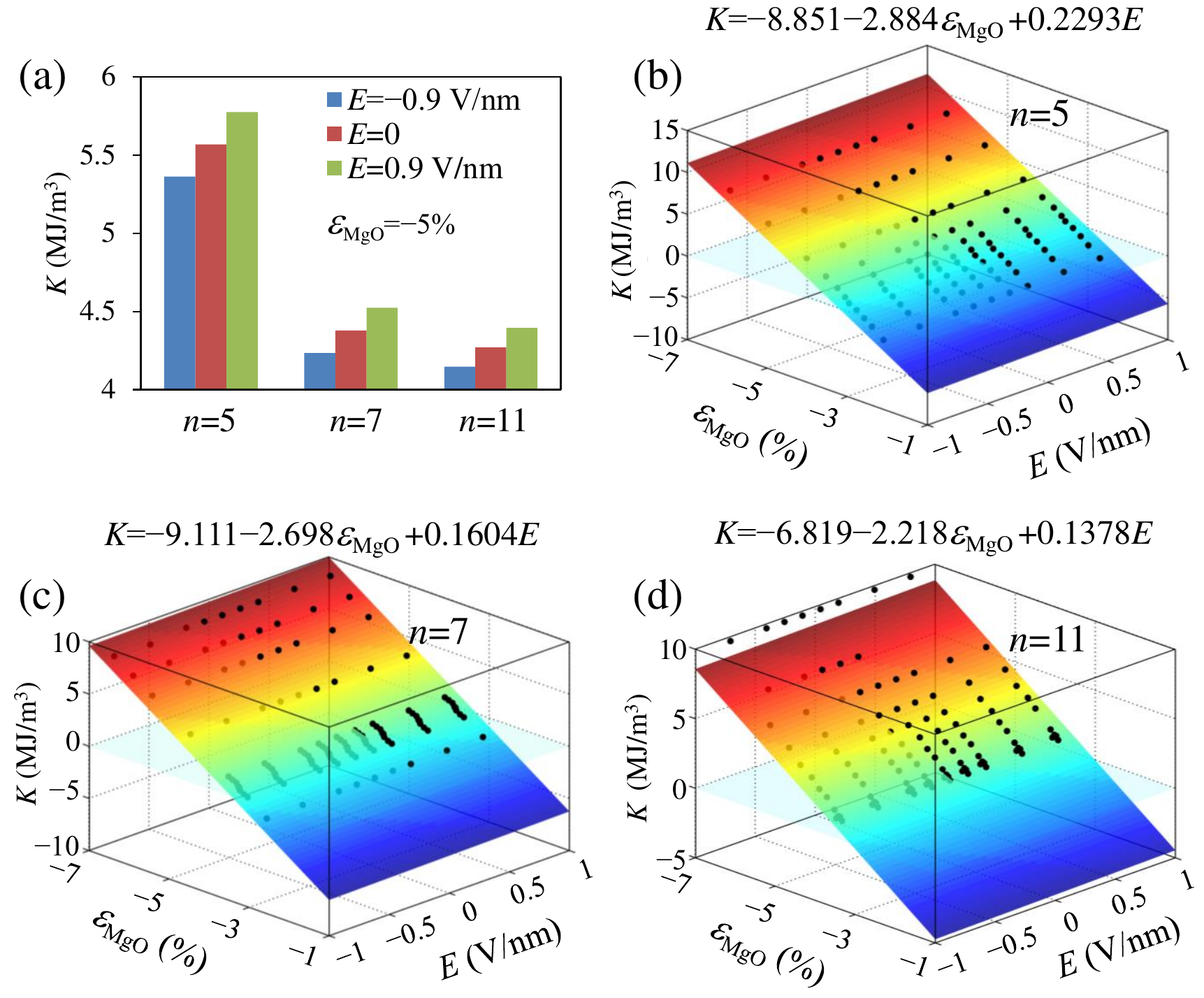}
\caption{(a) $K$ for different layer number $n$ of FePt. $K$ as functions of $\varepsilon_\text{MgO}$ and $E$ when (b) $n=5$, (c) $n=7$, and (d) $n=11$. The equations in (b)-(d) are obtained by linear fitting.}
\label{f3}
\end{figure}

Figure \ref{f2} presents the first-principles results when $\varepsilon_\text{MgO}=-3.26\%$ and $n=5$. 
From the electrostatic potential distribution obtained from first-principles calculation in Fig. {\ref{f2}}(a),
we can estimate the external electric field as 0.9 V/nm. Under this electric field, the charge density around Pt and Fe atoms which are close to the interface is evidently decreased. The effect of electric field extends to the first two layers of FePt next to the interface. The charge change will affect the $d$-orbital hybridization and thus the MAE, as shown the orbital-resolved MAE in Fig. S1 (in supplementary information, SI). The atom-resolved MAE analysis in Fig. \ref{f2}(d) indicates the dominative role of the interfacial Pt (Pt1) in the perpendicular magnetic anisotropy. The electric field induced change in MAE of Pt1 controls the MAE change of the system. Orbital-resolved MAE analysis in Fig. S1 (in SI) further confirms the largest MAE contribution of the hybridization between $d_{yz}$ and $d_{z^2}$ of Pt1. MAE of the system is shown to be linearly dependent on the strength of the electric fields, as presented in Fig. \ref{f2}(c). A giant modulation of MAE by applying electric fields is observed, with a amplitude as high as 322 fJ/Vm.

We then calculate saturation magnetization ($M_s$, total magnetic moment divided by FePt volume) and $K$ (MAE divided by FePt volume) under different $\varepsilon_\text{MgO}$, $E$ and $n$, as presented in Fig. S2 (in SI) and Fig. \ref{f3}, respectively. $M_s$ is found to be not significantly affected by $E$, but depend more on $\varepsilon_\text{MgO}$ and $n$ (Fig. S2 in SI). It can be seen from Fig. \ref{f3} that $K$ varies linearly with respect to both $\varepsilon_\text{MgO}$ and $n$, as indicated by the fitted linear functions therein. $K$ is more sensitive to $\varepsilon_\text{MgO}$ and can be altered from $\sim 10$ to $\sim -5$ MJ/m$^3$ when $\varepsilon_\text{MgO}$ is increased from $-7\%$ to $-1\%$. The strong dependence of $K$ on $\varepsilon_\text{MgO}$ intrigues extensive studies of the strain-mediated magnetization switching. Positive electric field results in an increment in $K$. The charge-mediated dependence of $K$ on $E$ provides the feasibility to switch the magnetization. For example, using the model in Fig. \ref{f1}(b), we find that through the charge-mediated $K$ change, the electric field can be used to tailor the magnetic hysteresis (Fig. S3 in SI).

\begin{figure}[t]
\centering
\includegraphics[width=8.6cm]{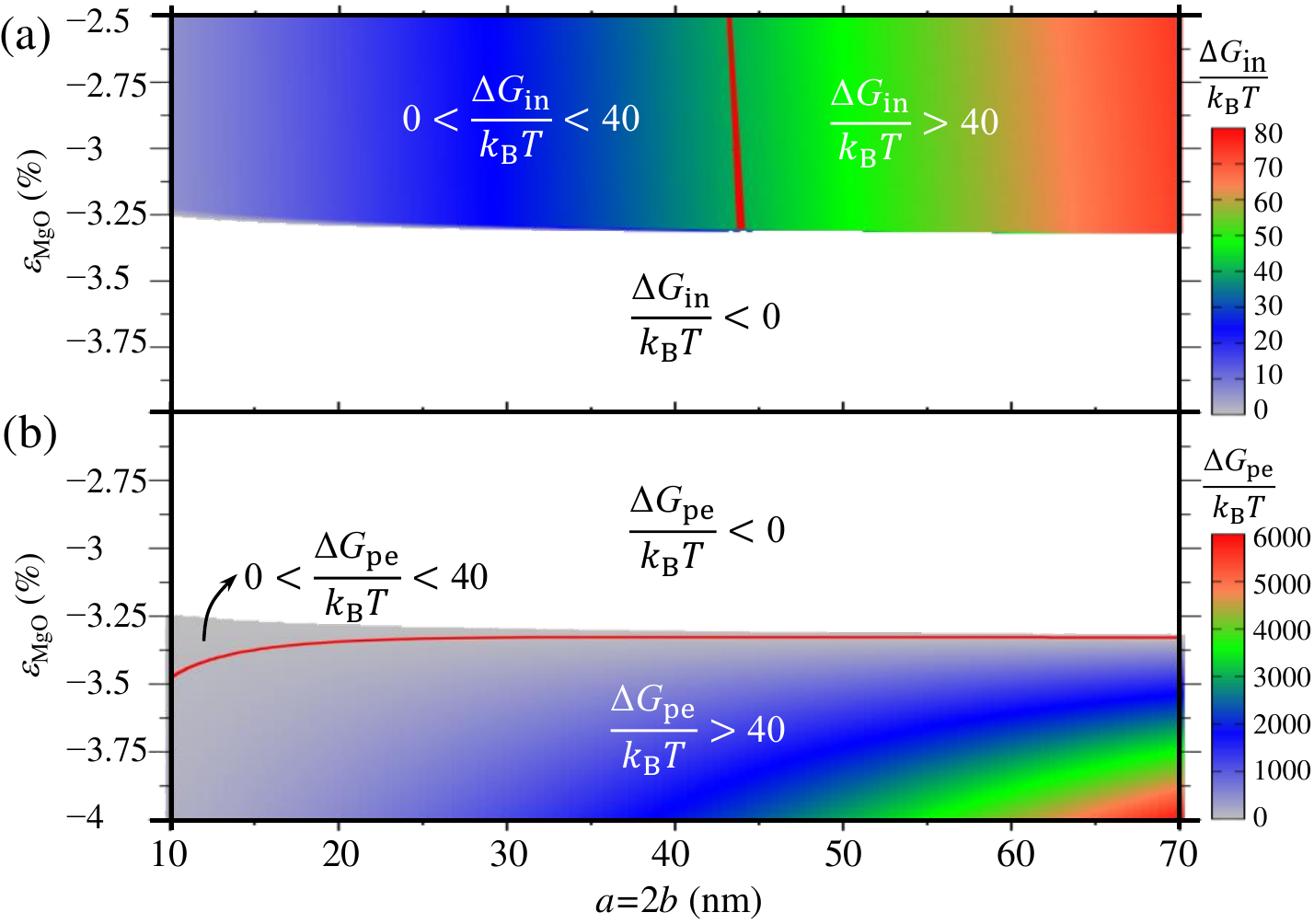}
\caption{Energy barrier of the initial equilibrium magnetization ($E=0$) with $n=11$: (a) initial magnetization along the in-plane $x$ direction; (b) initial magnetization along the perpendicular $z$ direction. $k_\text{B}$ is Boltzmann constant and $T=300$ K.}
\label{f4}
\end{figure}

From the perspective of practical applications of magnetic states in data storage, the energy barrier is usually required to exceed 40 $k_\text{B}T$ to make an initial equilibrium magnetization sufficiently stable. For this requirement, in the case of no electric field ($E=0$), we calculate the energy barrier ($\Delta G$) of initial magnetization stabilized in both in-plane $x$ and perpendicular $z$ directions, in order to check the thermal stability of the initial magnetization. 
As for the initial equilibrium magnetization state along the in-plane $x$ direction, the energy barrier is calculated by the energy difference between $x$ and $y/z$ directions when $E=0$, i.e. $\Delta G_\text{in}=\frac{1}{2}\mu_0 M_\text{s}^2 V[\text{min}(N_y, N_z-\frac{2K_{E=0}}{\mu_0 M_\text{s}^2})-N_x]$, in which the volume $V=\pi abt$ and $N_x$, $N_y$, and $N_z$ is the demagnetization factor along $x$, $y$, and $z$ direction, respectively. Similarly, in the case of initial magnetization along the perpendicular $z$ direction, the energy barrier can be calculated as $\Delta G_\text{pe}=\frac{1}{2}\mu_0 M_\text{s}^2 V[N_x-(N_z-\frac{2K_{E=0}}{\mu_0 M_\text{s}^2})]$. For simplicity, the elliptical geometry with $a=2b$ is considered. From these equations, we can determine the region in which $\Delta G$ is higher than 40 $k_\text{B}T$ ($T=300$ K), as shown by the contour plots in Fig. {\ref{f4}} ($n=11$) and Fig. S4 (in SI) ($n=5,7,9$). It can be seen from Fig. {\ref{f4}}(a) that for an in-plane switching with the initial magnetization along $x$ axis, suitable $\varepsilon_\text{MgO}$ larger than $\sim -3.313\%$ and $a$ larger than $\sim44$ nm are necessary for a energy barrier above 40 $k_\text{B}T$. The critical value of $a$ for an in-plane switching increases with the decrease of $n$ (Fig. S4 in SI). In contrast, a energy barrier over 40 $k_\text{B}T$ for a perpendicular switching with the initial magnetization along $z$ axis requires $\varepsilon_\text{MgO}$ smaller than $\sim -3.313\%$, but its dependence on $a$ is much weaker, as shown in Fig. {\ref{f4}}(b). Based on the results in Fig. {\ref{f4}}, the geometry parameter $a=2b=46$ nm is chosen for the following study of case $n=11$. This geometry makes it possible that the energy barrier of the initial equilibrium magnetization in both in-plane and perpendicular switching cases is beyond 40 $k_\text{B}T$ for practical applications.

\begin{figure}[t]
\centering
\includegraphics[width=8.6cm]{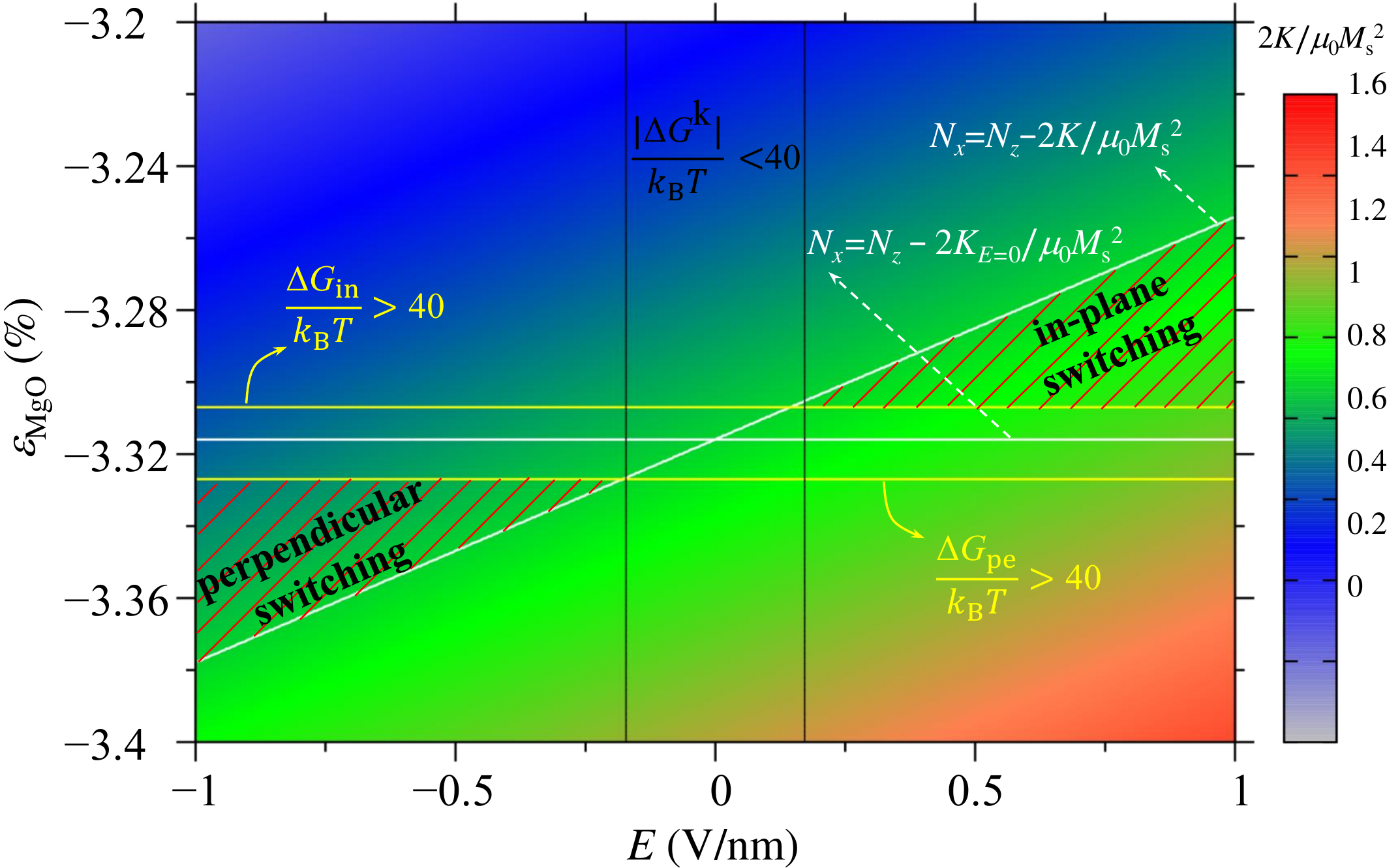}
\caption{Necessary conditions for the in-plane and perpendicular 180$^\circ$ switching ($n=11$, $a=2b=46$ nm): switching criteria $N_z-2K_{E=0}/\mu_0M_s^2>N_x>N_z-2K_{E>0}/\mu_0M_s^2$ or $N_z-2K_{E<0}/\mu_0M_s^2>N_x>N_z-2K_{E=0}/\mu_0M_s^2$, $E$ induced energy change $\vert\Delta G^\text{k}\vert$ beyond 40 $k_\text{B}T$, initial energy barrier $\Delta G_\text{in}$ $(\Delta G_\text{pe})$ over 40 $k_\text{B}T$.}
\label{f5}
\end{figure}

Heretofore we have figure out the strain and geometry conditions for a stable equilibrium magnetization ($E=0$) with energy barrier higher than 40 $k_\text{B}T$ from the viewpoint of practical applications. However, for the voltage-driven 180$^\circ$ switching, the $E$ induced energy change ($\vert\Delta G^\text{k}\vert$) and magnetization dynamics should be deliberated. Through the total energy analysis (Eq. {\ref{eq1}} in Methods), we firstly clarify how large $\varepsilon_\text{MgO}$ and $E$ make $\vert\Delta G^\text{k}\vert$ beyond 40 $k_\text{B}T$ and the 180$^\circ$ switching possible.
$E$ induced energy change can be easily calculated as $\vert\Delta G^\text{k}\vert=(K-K_{E=0})V$.
In the case of an in-plane 180$^\circ$ switching, the magnetic easy axis is along the semimajor axis $x$ when there is no electric field. The in-plane 180$^\circ$ switching means the magnetization switching from the positive semimajor axis to the negative one, or vice versa. These two equivalent in-plane easy axes guarantee a non-volatile effect, i.e. when the 180$^\circ$ switching is achieved the associated magnetization state will be kept even after the removal of the electric field. The necessary conditions for the 180$^\circ$ switching imply two criteria. Firstly, $K$ is small enough to ensure a magnetic easy plane when there is no electric field, i.e. $N_x<N_z-2K_{E=0}/\mu_0M_s^2$ (Eq {\ref{eq1}} in Methods). Secondly, after applying a positive electric field, $K$ is increased to be high enough to achieve a perpendicular easy axis, i.e. $N_x>N_z-2K_{E>0}/\mu_0M_s^2$. Therefore, the in-plane switching requires the criterion $N_z-2K_{E=0}/\mu_0M_s^2>N_x>N_z-2K_{E>0}/\mu_0M_s^2$.
On the contrary, the perpendicular 180$^\circ$ switching denotes the magnetization switching from the positive perpendicular axis $z$ to the negative one, or vice versa. Two criteria required by the necessary conditions for the perpendicular 180$^\circ$ switching are opposite to those for the in-plane 180$^\circ$ switching. On the one hand, without electric field $K$ should be large enough to ensure a perpendicular magnetic easy axis. On the other hand, after a negative electric field is applied, $K$ should be decreased to be small enough to ensure an in-plane easy axis. So the criterion $N_z-2K_{E<0}/\mu_0M_s^2>N_x>N_z-2K_{E=0}/\mu_0M_s^2$ should be fulfilled for the perpendicular switching.
Combing these two switching criteria, the relation $\vert\Delta G^\text{k}\vert>40$ $k_\text{B}T$, and the initial energy barrier $\Delta G_\text{in}$ $(\Delta G_\text{pe})>40$ $k_\text{B}T$ in Fig. {\ref{f4}}, we can estimate the necessary conditions for the possible 180$^\circ$ switching, as shown in Fig. {\ref{f5}}. It can be found that in the triangular regions within a small range of $\varepsilon_\text{MgO}$, a positive and negative $E$ potentiates an in-plane and perpendicular 180$^\circ$ switching, respectively. In a given system with specified layer number and $\varepsilon_\text{MgO}$, only one switching mode is possible. There is no overlapped region between in-plane and perpendicular switching modes.

\begin{figure}[t]
\centering
\includegraphics[width=8.6cm]{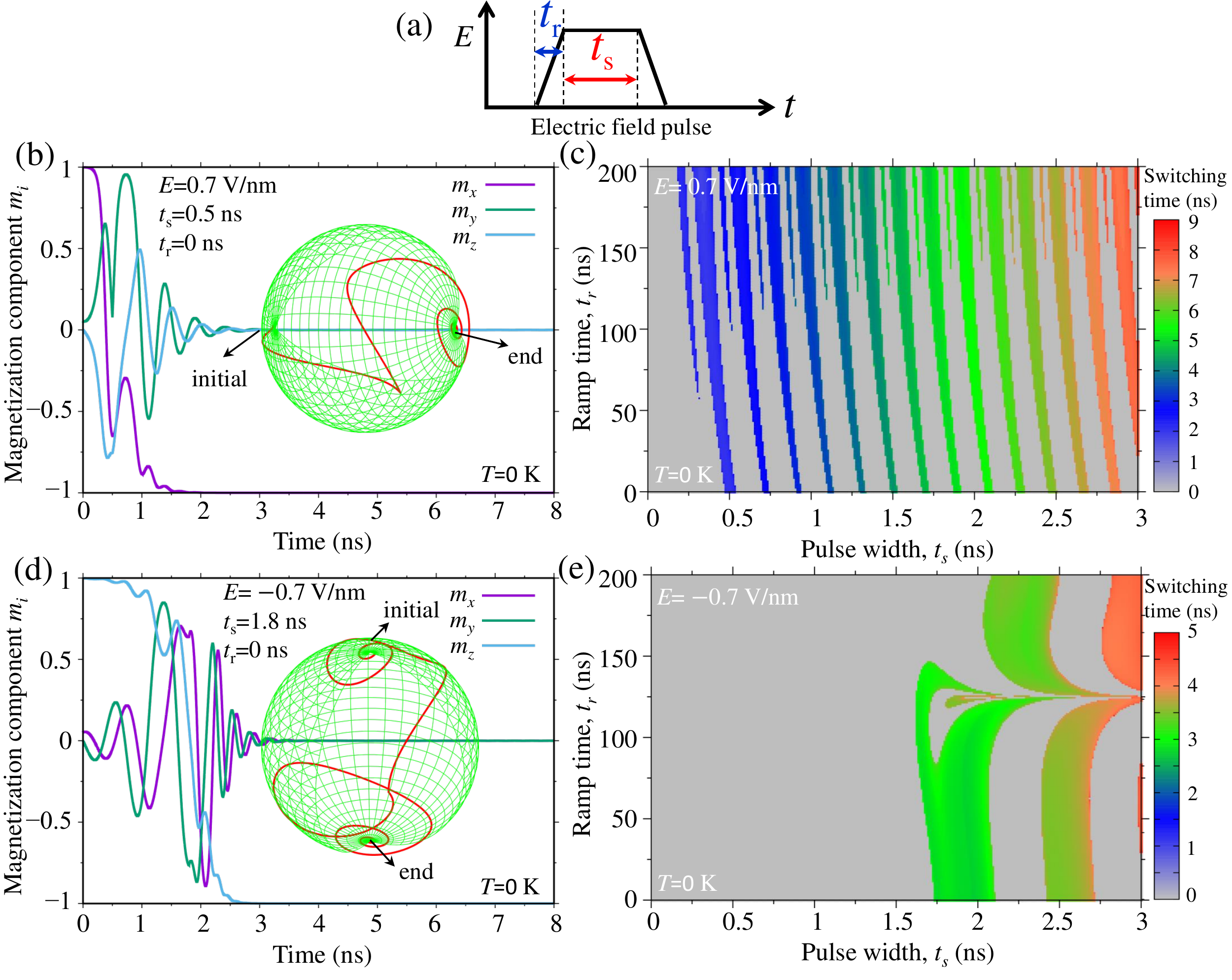}
\caption{(a) Electric field pulse with ramp time ($t_r$) and pulse width ($t_s$). Spatial and temporal trajectories of magnetization switching for (b) in-plane and (d) perpendicular 180$^\circ$ switching. Switching time as functions of $t_r$ and $t_s$ for (c) in-plane and (e) perpendicular 180$^\circ$ switching. The grey region or a switching time of zero in (c) and (e) means that the 180$^\circ$ switching fails. Temperature is set as 0 K.}
\label{f6}
\end{figure}

It should be noted that Fig. \ref{f5} only presents the necessary rather than the sufficient conditions for a 180$^\circ$ switching. If the electric field is always kept until the magnetization reaches the equilibrium state, only a 90$^\circ$ switching is achieved. After removing the electric field, in the viewpoint of static analysis the magnetization possesses the equal possibility to revert back or go forwards. Only by using the 3D precessional switching dynamics of magnetization,\cite{12,13,14,17,18,23,24,26} a deterministic 180$^\circ$ switching can be achieved when the pulse width of the electric field is controlled. In the following, we will focus on the switching dynamics in the case of $n=11$ and $\varepsilon_\text{MgO}=-3.29\%$ for an in-plane 180$^\circ$ switching and $\varepsilon_\text{MgO}=-3.34\%$ for a perpendicular 180$^\circ$ switching. An electric field pulse with ramp time ($t_r$) and pulse width ($t_s$) shown in Fig. \ref{f6}(a) is applied to provoke switching. As a first step, we studied the typical switching dynamics at 0 K, as presented in Fig. \ref{f6}(b)-(e). The spatial and temporal magnetization trajectories in Fig. \ref{f6}(b) and (d) indicate an in-plane and perpendicular 180$^\circ$ switching, respectively. It can be seen that the precessional dynamics after the removal of $E$ at $t_s$ make the magnetization component towards $-1$ possible. The control of $t_r$ and $t_s$ is critical for realizing the 180$^\circ$ switching, as shown in Fig. \ref{f6}(c) and (e). It can be found that in both the in-plane and perpendicular 180$^\circ$ switching, a fast switching within $\sim$3 ns can be attained. The ramp time plays a role in determining whether the 180$^\circ$ switching fails or succeeds, but has little effects on the switching time which is more influenced by the pulse width. The in-plane 180$^\circ$ switching allows a much wider range of $E$ pulse width than the perpendicular one. These switching dynamic features are consistent with those from the literature in which the temperature-induced thermal fluctuations are ignored.

\begin{figure}[b]
\centering
\includegraphics[width=7.6cm]{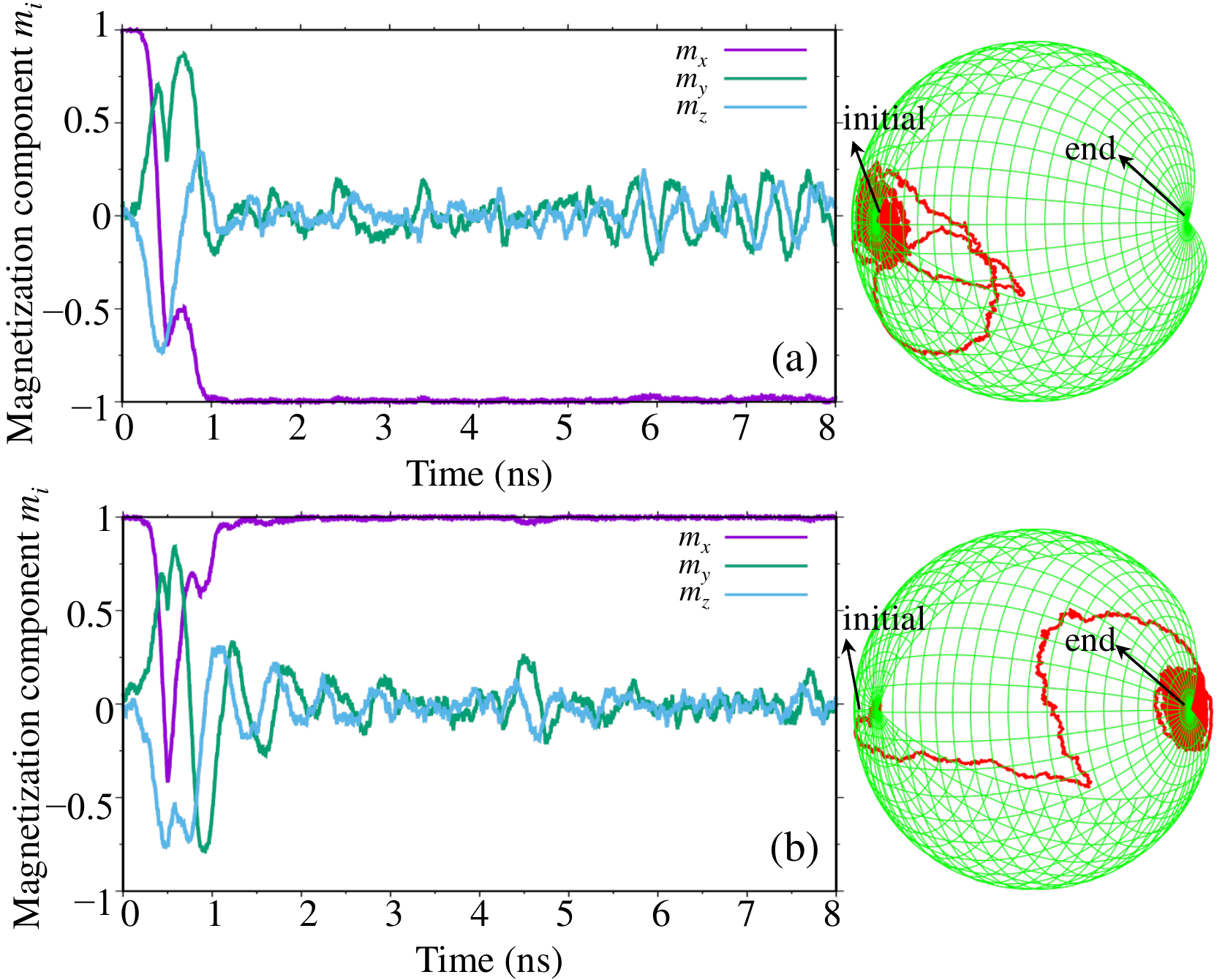}
\caption{Two individual trajectories at $T=300$ K for an in-plane 180$^\circ$ switching ($E=0.7$ V/nm, $t_s=$0.5 ns, $t_r$=0): (a) 180$^\circ$ switching succeeds; (b) 180$^\circ$ switching fails.}
\label{f7}
\end{figure}

\begin{figure}[t]
\centering
\includegraphics[width=8.6cm]{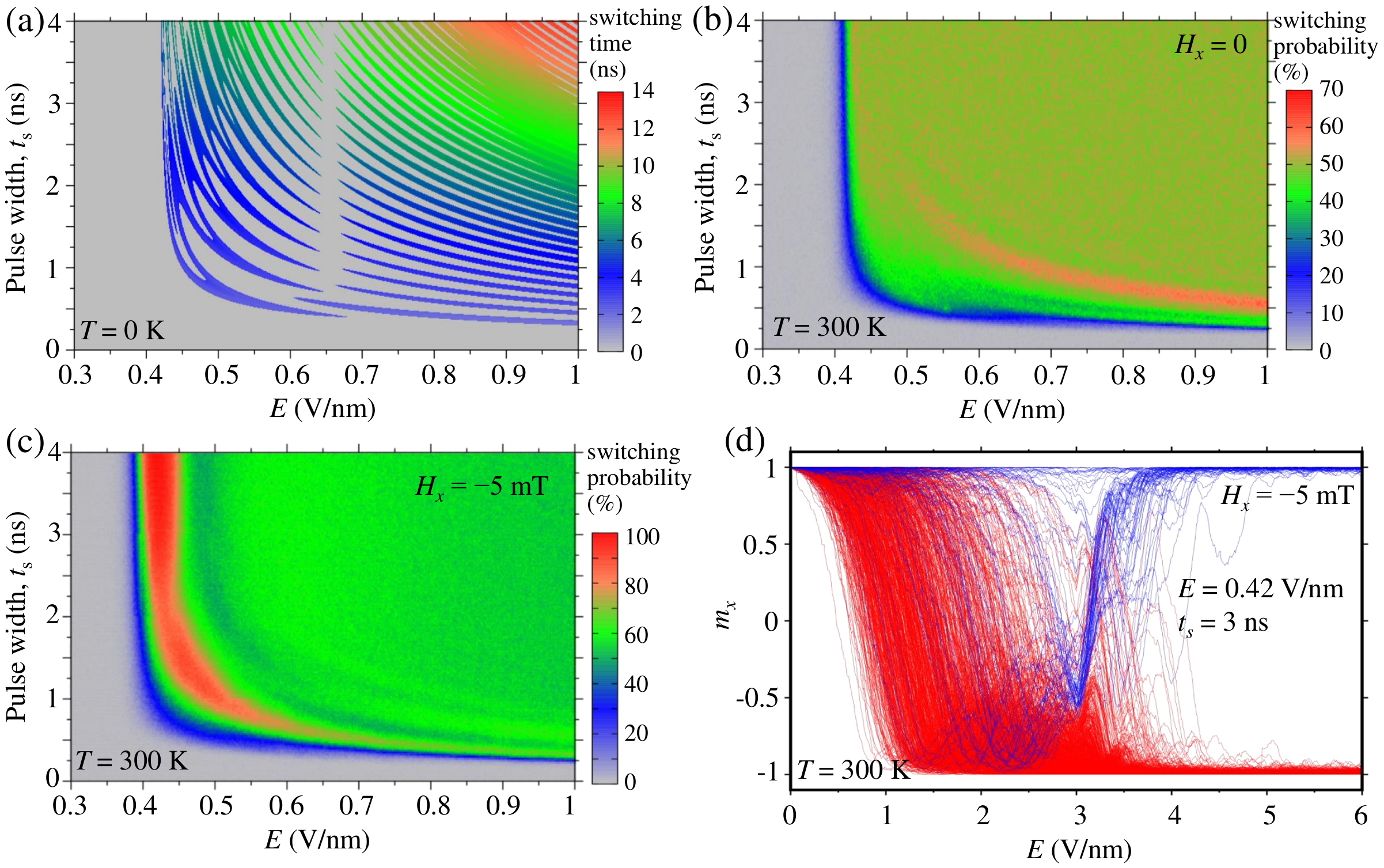}
\caption{In-plane 180$^\circ$ switching. (a) Switching time at 0 K. Switching probability at 300 K: (b) no bias field and (c) bias field $H_x=-5$ mT. (d) 1,000 trajectories with a switching probability of $\sim93.2\%$ at 300 K. $t_r=0$.}
\label{f8}
\end{figure}

\begin{figure}[t]
\centering
\includegraphics[width=8.6cm]{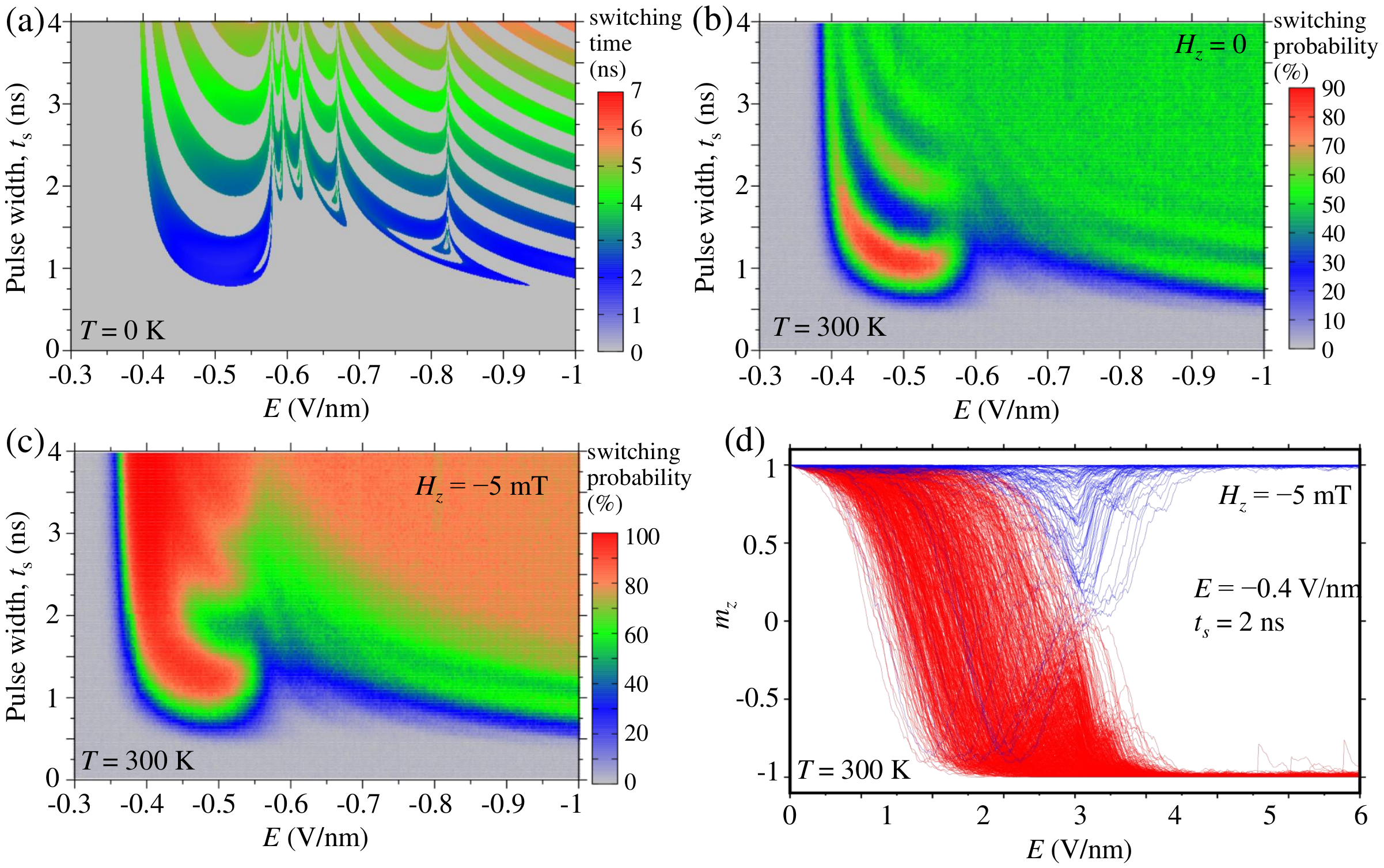}
\caption{Perpendicular 180$^\circ$ switching. (a) Switching time at 0 K. Switching probability at 300 K: (b) no bias field and (c) bias field $H_z=-5$ mT. (d) 1,000 trajectories with a switching probability of $\sim91.5\%$ at 300 K. $t_r=0$.}
\label{f9}
\end{figure}

However, if the temperature-induced thermal fluctuations is considered, the switching dynamics can be intrinsically altered. Thereby the thermal fluctuation of torque should be considered, as explained in the Methods.
Here we do not perform the temperature-dependent first-principles calculations, but focus on the temperature effect on switching dynamics, i.e. magnetization dynamics including temperature-induced thermal fluctuations. A finite temperature study should be carried out further so that temperature effects can be included in both the parameter prediction by first-principles calculations and the simulation of magnetization dynamics.
Even in the equilibrium states without electric field, the magnetization is not exactly aligned along the easy axis. The finite temperature effect makes the magnetization fluctuate within several degrees around the easy axis (Fig. S5 in SI). Figure \ref{f7} presents two individual magnetization trajectories at $T=300$ K, with other conditions the same as those in Fig. \ref{f6}(a). It is clear that Fig. \ref{f6}(a) without temperature effect shows a deterministic 180$^\circ$ switching. But if $T=300$ K is taken into account, the 180$^\circ$ switching can either succeed (Fig. \ref{f7}(a)) or fail (Fig. \ref{f7}(b)). It indicates that the temperature effect disturbs the switching behavior and makes the 180$^\circ$ switching as probability events. The deterministic switching demonstrated at 0 K may be not really deterministic at finite temperature, making the previous studies of the 180$^\circ$ switching at 0 K to be re-examined.

Figs. \ref{f8} and \ref{f9} compare the switching behaviors at 0 K with those at 300 K. It can be found from Fig. \ref{f8}(a) and Fig. \ref{f9}(a) that at 0 K, the 180$^\circ$ switching is deterministic and possesses a characteristic switching time. Large electric fields and small pulse width favor a fast switching, with a minimum switching time around 2 ns. When the temperature is involved in the switching dynamics, statistical methodology should be applied to get the 180$^\circ$ switching probability.

As shown in Fig. \ref{f8}(b), \ref{f8}(c) and Fig. \ref{f9}(b), \ref{f9}(c), at 300 K the switching is not deterministic and a switching probability (the percentage of successful 180$^\circ$ switching) or error probability (the percentage of unsuccessful 180$^\circ$ switching) is obtained. For the in-plane 180$^\circ$ switching, Fig. \ref{f8}(b) shows that the room temperature makes the 180$^\circ$ switching with a maximum probability of $~70\%$. If we apply a small bias magnetic field $H_x=-5$ mT which is the strength of a typical refrigerator magnet, a switching probability above $90\%$ can be achieved (Fig. \ref{f8}(c)). For the perpendicular 180$^\circ$ switching in Fig. \ref{f9}, we observe the similar behavior. The magnitude and the pulse width of the electric fields should be carefully designed for a low error probability at room temperature. The region with probability above $\sim 90\%$ is much wider in Fig. \ref{f9}(c) than that in Fig. \ref{f8}(c), indicating more controllability in the perpendicular 180$^\circ$ switching.

For an estimation of switching time at room temperature, as an example we present 1,000 switching trajectories with relatively low error probability for in-plane switching ($E=0.42$ V/nm and $t_s=3$ ns, Fig. \ref{f8}(d)) and perpendicular switching ($E=-0.4$ V/nm and $t_s=2$ ns, Fig. \ref{f9}(d)). The switching time is found to be around 4 ns. The switching probability in Fig. \ref{f8}(d) and Fig. \ref{f9}(d) is $\sim93.2\%$ (error probability $\sim6.8\%$) and $\sim91.5\%$ (error probability $\sim8.5\%$), respectively. Undeniably, decreasing the error probability as much as possible is desirable. However, the achieved switching probability around $90\%$ here is still reasonable or may be adequate for memory applications where different on-chip error detection and correction schemes exist.

In summary, the voltage-driven charge-mediated perpendicular and in-plane 180$^\circ$ magnetization switching at 0 K and room temperature is studied by using a multiscale theoretical framework which combines first-principles calculations and temperature-dependent magnetization dynamics. For the epitaxial metal-magnet-insulator (Pt/FePt/MgO) hetero-nanostructure as the model system, it is found from first-principles calculations that the interfacial charges induced by electric fields induce a giant modulation of MAE of the nanomagnet. From the temperature-dependent magnetization dynamics using first-principles results, it is found that both in-plane and perpendicular 180$^\circ$ magnetization switching is possible in the case of suitable epitaxial strain, $E$ pulse width, and $E$ ramp rate. But the temperature effect disturbs the switching behavior and makes the 180$^\circ$ switching as probability events. The $E$ magnitude and pulse width should be carefully designed for a low-error-probability 180$^\circ$ switching at room temperature. Statistical analysis indicates that a fast (around 4 ns) 180$^\circ$ switching of low error probability can be achieved at room temperature. This work not only demonstrates a charge-mediated way for controlling magnetization by voltage, but also inspires the rational design of miniaturized nanoscale spintronic devices where temperature induced thermal fluctuation plays a critical role.

\section*{Methods}
The MAE and $M_s$ of the system in Fig. \ref{f1}(a) are mainly originated from the $L1_0$ ordered FePt layer. The supercell is constructed along the (001) direction, containing $n$-layer FePt on top of four-layer MgO followed by three-layer Pt and a 15-\AA-thick vacuum layer. The FePt layer number $n$ is chosen to be 5, 7, 9, and 11.
The electric field is imposed by the dipole layer method,{\cite{53}} with the dipole placed in the middle of the vacuum region.
The first-principles calculations were carried out within the density functional theory and the framework of the projector augmented-wave formalism as implemented in the Vienna \textit{ab initio} simulation package (VASP).\cite{54} The Perdew-Burke-Ernzerhof (PBE) exchange-correlation functional in the generalized gradient approximation (GGA) was employed. An energy cutoff of 500 eV and a Monkhorst-Pack $k$-mesh $31\times31\times1$, at which a good convergence of MAE was achieved, were utilized.
At a certain $\varepsilon_\text{MgO}$, the in-plane lattice parameter of the supercell is fixed to be that of the strained MgO during the relaxation and the atomic positions in the $z$ direction are relaxed.
The convergence criteria for the structure relaxation at different $\varepsilon_\text{MgO}$ were set as $10^{-6}$ eV and 2 meV/\AA ~ for the energies and forces, respectively. By using the self-consistent charge density, non-self-consistent calculations with spin-orbit coupling were performed to get the total energy as a function of the orientation of the quantization axis.
$K$ was evaluated as the difference of the total energy per unit FePt volume when the magnetization was along (100)/(010) ($x/y$) and (001) ($z$) directions. Positive and negative $K$ indicates perpendicular and in-plane magnetic anisotropy, respectively.

The material system for micromagnetic dynamics analysis is shown in Fig. \ref{f1}(b). The FePt nanomagnet is an elliptical cylinder with height $t(n)$, semimajor axis $a=46$ nm, and semiminor axis $b=23$ nm. For such a small size, a single domain exists and two angles ($\theta,\phi$) are used to describe the magnetization state.
The single-domain state of FePt nanomagnet with magnetic properties from first principles is confirmed by our micromagnetic simulations (Fig. S6 in SI). It should be noted that the electric field induced $K$ change is mainly localized around the interface. This interface effect will be much weaker for the thick film. Here, the FePt nanomagnet is only several atomic layer thick. Since the magnetization behaves coherently (Fig. S6 in SI), we represent all the magnetic moments in these atomic layers by a macro spin with the average saturation magnetization $M_s$ and $K$ from first-principles calculations. The similar idea has also been used by previous work.{\cite{31,37}} Nevertheless, a more accurate way by atomistic spin simulations which treat each atomic spin separately should be attempted as the next-step work. In this way, the total energy of the FePt elliptical cylinder can be given as the summation of the demagnetization energy and the magnetocrystalline anisotropy energy, i.e.
\begin{equation}
\begin{split}
E_t=\frac{1}{2}\mu_0M_s^2 \Bigg[& N_x\sin^2\theta\cos^2\phi+N_y\sin^2\theta\sin^2\phi+  \\
&  \left(N_z-\frac{K}{\frac{1}{2}\mu_0M_s^2}\right)\cos^2\theta \Bigg]
\end{split}
\label{eq1}
\end{equation}
in which $N_x$, $N_y$, and $N_z$ is the demagnetization factor along $x$, $y$, and $z$ direction, respectively, and can be calculated as a function of the geometry size.\cite{55} Taking the temperature effect as thermal fluctuations,\cite{56} the temperature-dependent magnetization dynamics is governed by
\begin{equation}
\begin{split}
\dot{\theta}=& -\frac{\gamma_0}{M_s(1+\alpha^2)}\left(\alpha\frac{\partial E_t}{\partial \theta}+\frac{1}{\sin\theta}\frac{\partial E_t}{\partial \phi}\right) + \\
& \frac{1}{2\tau_N}\cot\theta + \frac{1}{\sqrt{\tau_N\Delta t}}P_1
\end{split}
\label{eq2}
\end{equation}

\begin{equation}
\begin{split}
\dot{\phi}= &-\frac{\gamma_0}{M_s(1+\alpha^2)\sin\theta}\left(\frac{\alpha}{\sin\theta}\frac{\partial E_t}{\partial \phi}-\frac{\partial E_t}{\partial \theta}\right) + \\
&  \frac{1}{\sin\theta\sqrt{\tau_N\Delta t}}P_2
\end{split}
\label{eq3}
\end{equation}
in which $\gamma_0$ is the gyromagnetic ratio constant, $\alpha=0.2$ is the damping coefficient of $L1_0$ FePt,\cite{57} $\Delta t=0.2$ ps is the time step, and $P_i$ ($i=$1, 2) is a stochastic process with Gaussian distribution, zero mean value, and completely uncorrelated property in time. The characteristic time $\tau_N$ is related to volume $V$ and temperature $T$ as $\tau_N^{-1}=2\alpha\gamma_0k_\text{B}T/[M_s(1+\alpha^2)V]$.
The method by using Eqs. {\ref{eq2}} and {\ref{eq3}} is verified by performing a benchmark test, in which only the thermal fluctuations are considered. In such a simple case, an initial magnetization ($m_x^0$, $m_y^0$, $m_z^0$) will evolve randomly and the associated average response can be obtained from the theoretical solution of Fokker-Planck equation as $\langle m_i \rangle=m_i^0 \text{exp}(-t/\tau_N)$. Our simulation results are found to agree well with the theoretical solution and the simulation results by MuMax3{\cite{58}}.

\section*{Acknowledgment} 
The financial supports from the German federal state of Hessen through its excellence programme LOEWE RESPONSE and the German Science Foundation (individual project Xu 121/7-1 and the project Xu 121/4-2 in the Forscher gruppe FOR1509) are appreciated. The authors also greatly acknowledge the access to the Lichtenberg High Performance Computer of TU Darmstadt.

\section*{AUTHOR CONTRIBUTIONS}
M. Y. performed the calculations, analyzed the results, and wrote the manuscript. H. Z. and B.-X. X. analyzed the results and supervised the project. All authors reviewed the manuscript.

\section*{COMPETING INTERESTS}
The authors declare no competing interests.

\newpage

\begin{figure*}[!h]
\renewcommand\thefigure{S1}
\centering
\includegraphics[width=15cm]{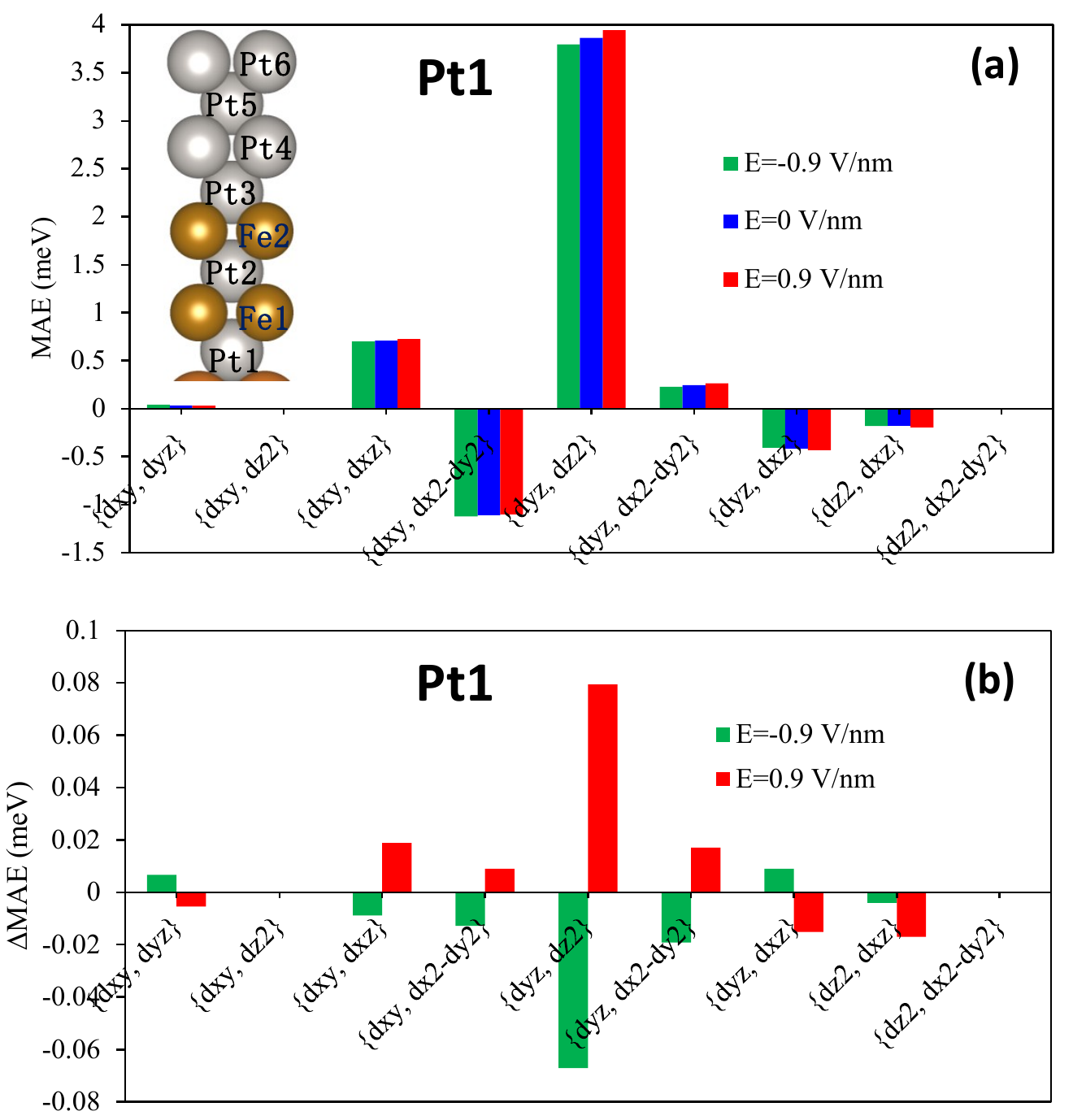}
\caption{(a) Orbital-resolved magnetic anisotropy energy (MAE) for the layer Pt1, when an electric field of 0.9, 0, and -0.9 V/nm is applied. (b) The induced changes in the orbital-resolved magnetic anisotropy energy ($\Delta$MAE) for the layer Pt1, when an electric field of 0.9 V/nm and $-0.9$ V/nm is applied. $n=5$ and $\varepsilon_\text{MgO}=-3.26\%$.}
\label{fS1}
\end{figure*}

\newpage
\begin{figure*}[!h]
\renewcommand\thefigure{S2}
\centering
\includegraphics[width=16cm]{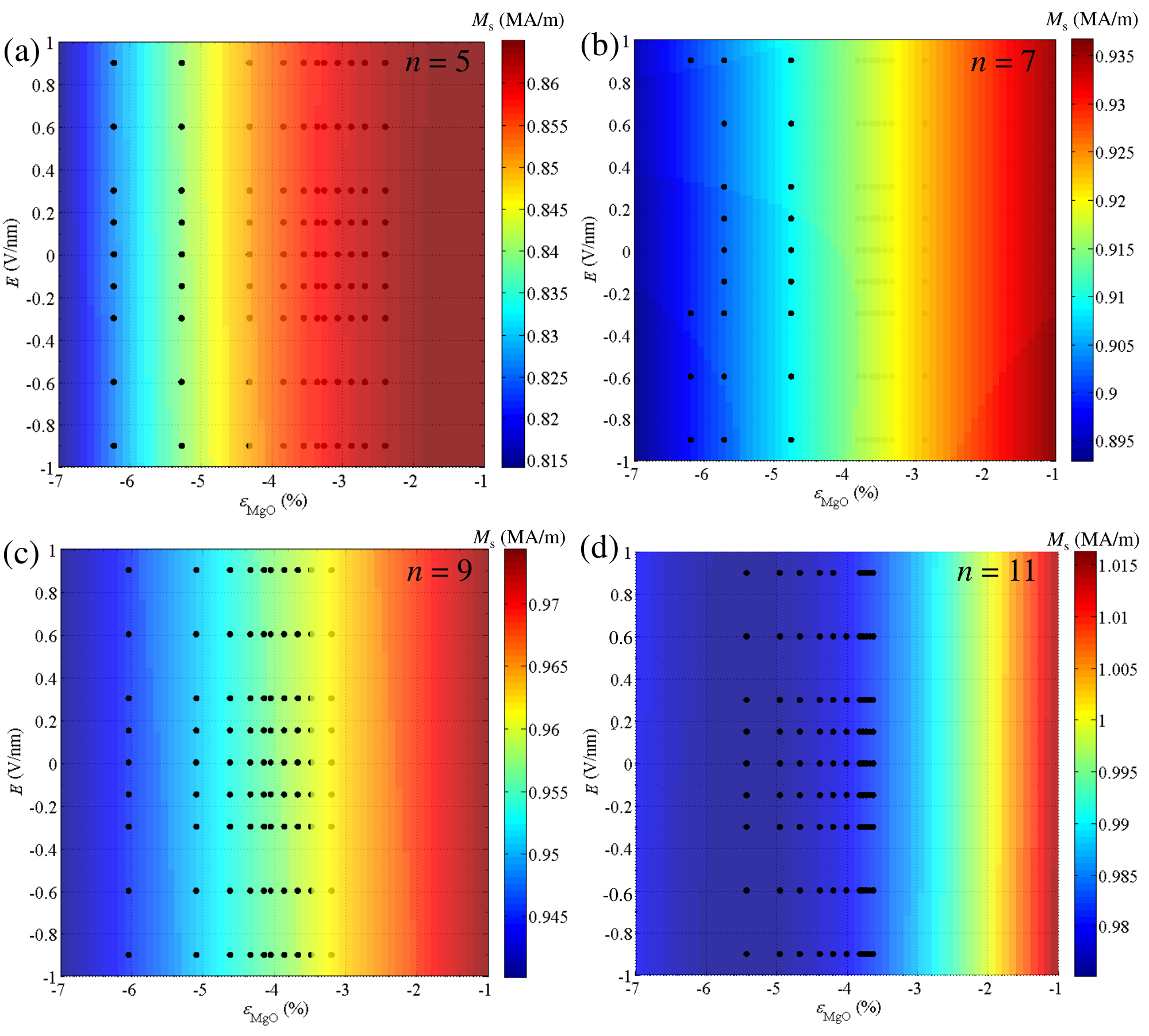}
\caption{First-principles results of $M_s$.}
\label{fS2}
\end{figure*}

\newpage
\begin{figure*}[!h]
\renewcommand\thefigure{S3}
\centering
\includegraphics[width=12cm]{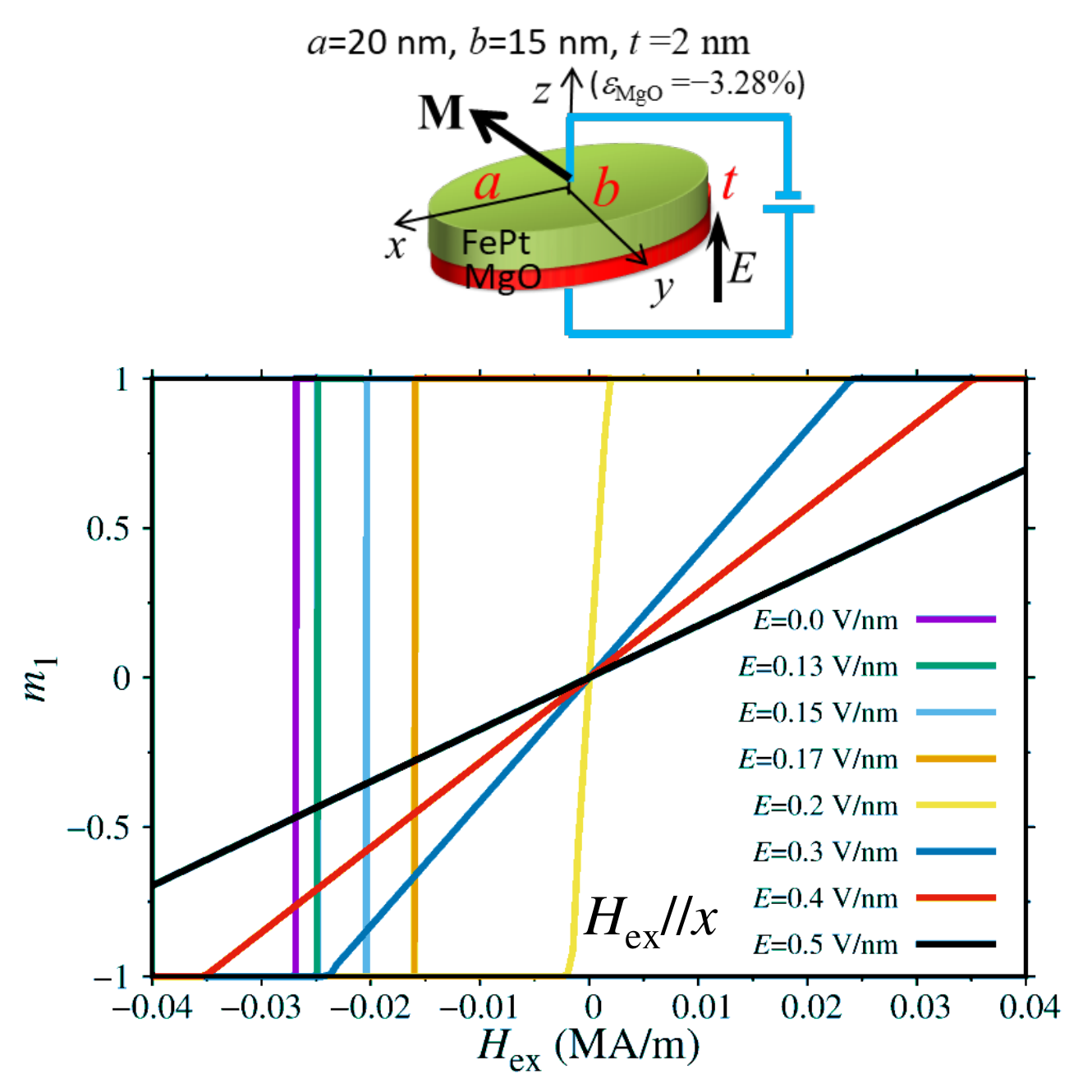}
\caption{Electric field dependent magnetic hysteresis.}
\label{fS3}
\end{figure*}

\newpage
\begin{figure*}[h!]
\renewcommand\thefigure{S4}
\centering
\includegraphics[width=16.5cm]{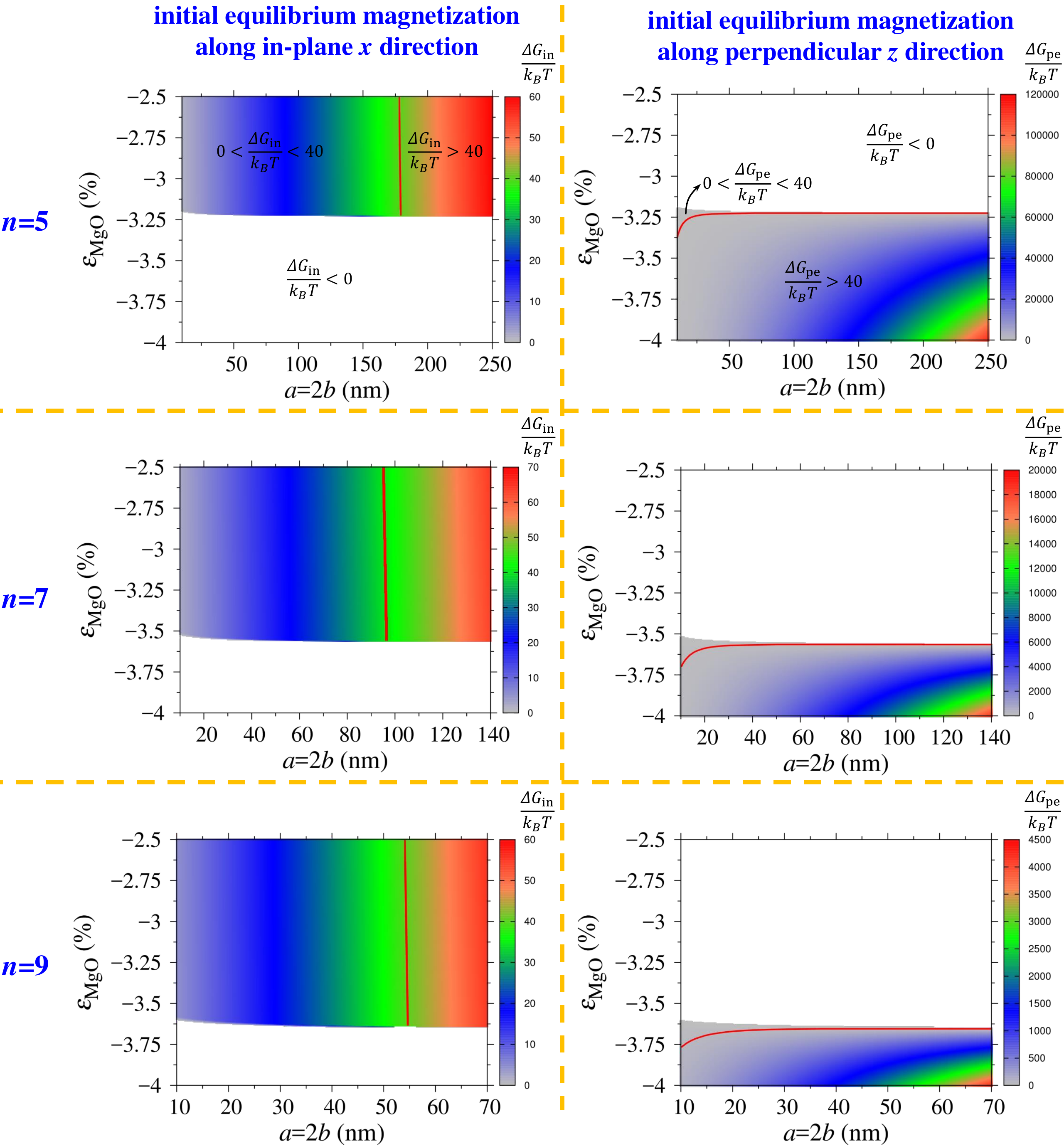}
\caption{Energy barrier of the initial equilibrium magnetization ($E=0$) with $n=5$, 7, and 9, as functions of elliptic geometry parameter $a=2b$ and $\varepsilon_\text{MgO}$. $k_\text{B}$ is Boltzmann constant and $T=300$ K.}
\label{fS4}
\end{figure*}

\newpage
\begin{figure*}[!h]
\renewcommand\thefigure{S5}
\centering
\includegraphics[width=12cm]{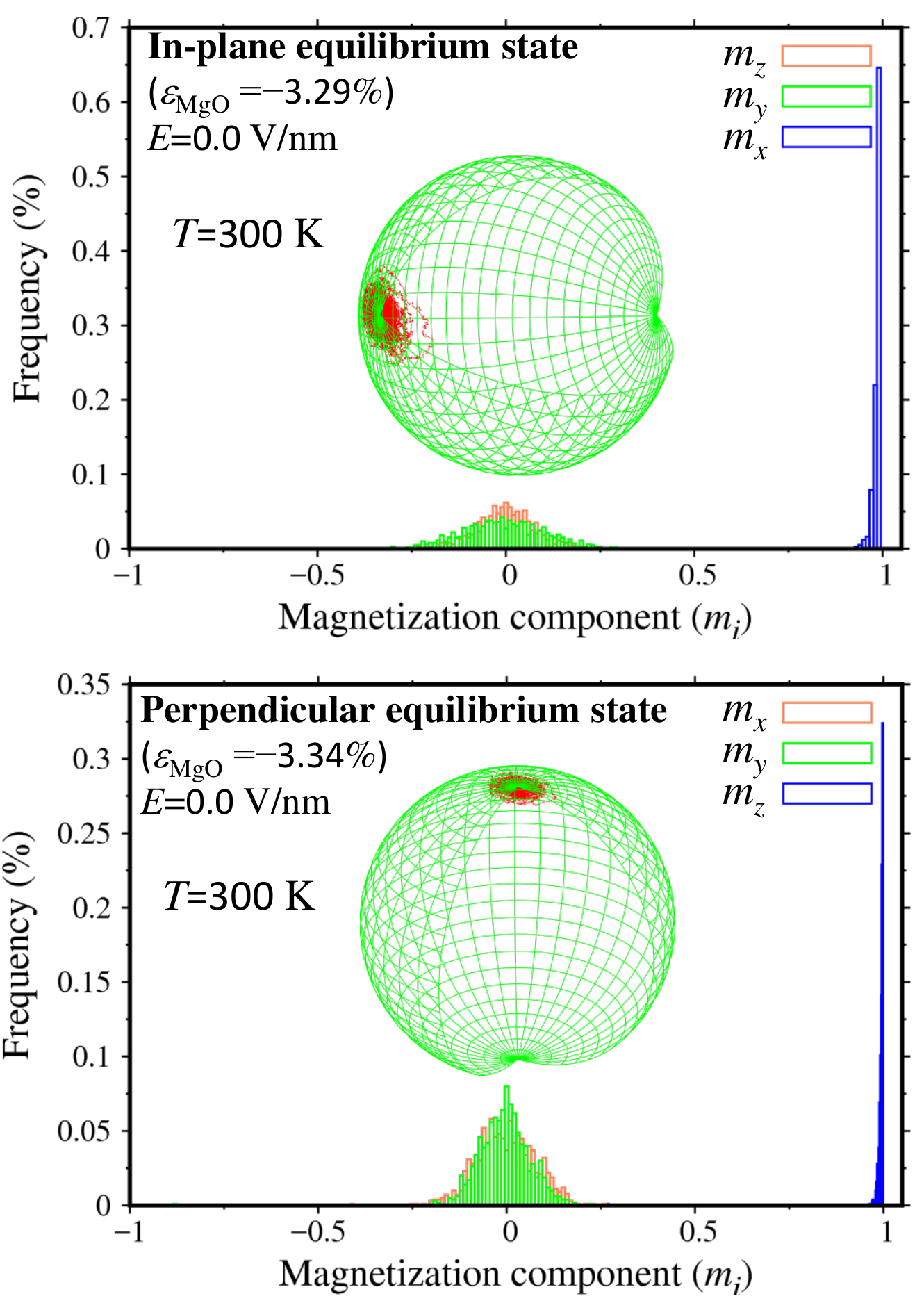}
\caption{Magnetization distribution in the equilibrium states at room temperature without electric field ($a=2b=46$ nm, $n=11$).}
\label{fS5}
\end{figure*}

\newpage
\begin{figure*}[!h]
\renewcommand\thefigure{S6}
\centering
\includegraphics[width=16cm]{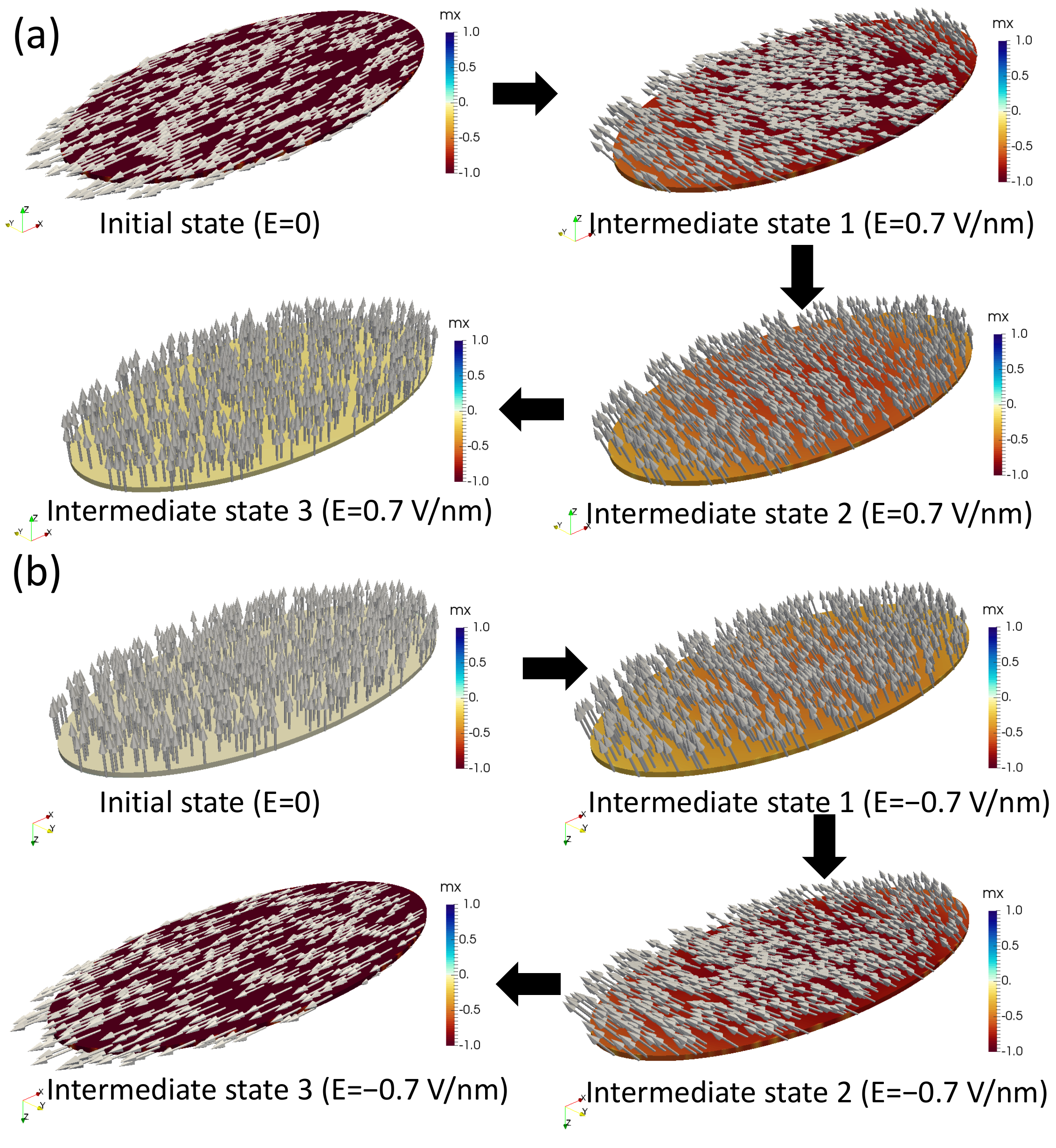}
\caption{ Electric field induced magnetic states predicted by micromagnetic simulations (OOMMF) when the initial state is (a) in-plane and (b) perpendicular single domain. Elliptical cylinder with height 2 nm, semimajor axis $a=46$ nm, and semiminor axis $b=23$ nm.}
\label{fS6}
\end{figure*}

\end{document}